\documentclass{elsart}

\usepackage{amsmath}
\usepackage{amssymb}
\usepackage{rotating}

\journal{Computer Physics Communications}

\DeclareMathOperator{\sech}{sech}

\newcounter{step}[section]
\renewcommand{\thestep}{\arabic{step}}
\newenvironment{step}[1]%
  { \stepcounter{step} \medskip \par \noindent \textsc{Step \thestep} (#1).}%
  {\medskip}
\renewcommand{\vec}[1]{\mathbf{#1}}
\begin{document}
\begin{frontmatter}

\title{Symbolic computation of hyperbolic tangent solutions for 
  nonlinear differential-difference equations\thanksref{titlefn}}

\thanks[titlefn]{This material is based upon research supported by 
  the National Science Foundation under Grants 
  Nos.\ CCR-9901929, DMS-9732069 and DMS-9912293.}

\author[CSM]{D.\ Baldwin}
\author[Wolfram]{\"{U}.\ G\"{o}kta\c{s}}
\author[CSM,SAfrica]{W.\ Hereman\corauthref{cor}}
\corauth[cor]{Corresponding author.}
\ead{whereman@mines.edu}
\ead[url]{http://www.mines.edu/fs\_home/whereman}

\address[CSM]{Department of Mathematical and Computer Sciences,
Colorado School of Mines, Golden, CO 80401-1887, U.S.A. }
\address[Wolfram]{Wolfram Research, Inc., 100 Trade Center Drive, 
Champaign, IL 61820, U.S.A.}
\address[SAfrica]{Department of Applied Mathematics, 
University of Stellenbosch, Private Bag X1, 7602 Matieland, South Africa}

\begin{abstract}
  A new algorithm is presented to find exact traveling wave solutions of 
  differential-difference equations in terms of tanh functions.  
  For systems with parameters, the algorithm determines the conditions on the 
  parameters so that the equations might admit polynomial solutions in tanh.  

  Examples illustrate the key steps of the algorithm.  
  Parallels are drawn through discussion and example to the tanh-method for 
  partial differential equations.

  The new algorithm is implemented in \emph{Mathematica}.  
  The package {\em DDESpecialSolutions.m\/} can be used to automatically 
  compute traveling wave solutions of nonlinear polynomial 
  differential-difference equations.  
  Use of the package, implementation issues, scope, and limitations of 
  the software are addressed.


  \noindent{\bf Program summary}

  \noindent{\em Title of program:\/} 
    DDESpecialSolutions.m \\
  \noindent{\em Catalogue identifier (supplied by the Publisher):\/} \\
  \noindent{\em Distribution format (supplied by the Program Library):\/} \\
  \noindent{\em Computers:\/} 
    Created using a PC, but can be run on UNIX and Apple machines \\
  \noindent{\em Operating systems under which the program has been tested:\/} 
    Windows 2000 and XP\\
  \noindent{\em Programming language used:\/} 
    Mathematica \\
  \noindent{\em Memory required to execute with typical data:\/} 
    9 MB \\
  \noindent{\em Number of processors used:\/} 
    1 \\
  \noindent{\em Has the code been vectorised or parallelized?:\/} 
    No \\
  \noindent{\em Number of bytes in distributed program, including test data, 
etc.:\/} 
    104 761

  \noindent{\em Nature of physical problem:\/} 
    The program computes exact solutions to differential-difference 
    equations in terms of the $\tanh$ function.  
    Such solutions describe particle vibrations in lattices, 
    currents in electrical networks, pulses in biological chains, etc.  \\
  \noindent{\em Method of solution:\/} 
    After the differential-difference equation is placed in a traveling 
    frame of reference, the coefficients of a candidate polynomial solution 
    in $\tanh$ are solved for.  
    The resulting solution is tested by substitution into the 
    original differential-difference equation.  \\
  \noindent{\em Restrictions on the complexity of the program:\/} 
    The system of differential-difference equations must be polynomial. 
    Solutions are polynomial in $\tanh.$ \\
  \noindent{\em Typical running time:\/} 
    The average run time of 16 cases 
    (such as Toda, Volterra, and Ablowitz-Ladik lattices) 
    is 0.228 seconds with a standard deviation of 0.165 seconds 
    on a 2.4GHz Pentium 4 with 512 MB RAM running Mathematica 4.1.  
    The running time may vary considerably, 
    depending on the complexity of the problem.  

\end{abstract}
\end{frontmatter}
%
\section{Introduction}
\label{sec:intro}
Since the work of Fermi, Pasta, and Ulam in the 1950s \cite{fermi}, 
differential-difference equations (DDEs) have been the focus of many 
nonlinear studies (for references see e.g.\ \cite{hickmanhereman03,teschl}).
There is renewed interest in DDEs, which can be used to model such physical 
phenomena as particle vibrations in lattices, currents in electrical networks, 
pulses in biological chains, etc.  
Unlike difference equations which are fully discretized, 
DDEs are semi-discretized with some (or all) of their spacial variables 
discretized while time is usually kept continuous.  
DDEs also play an important role in numerical simulations of nonlinear 
partial differential equations (PDEs),
queuing problems, and discretizations in solid state and quantum physics.

There is a vast body of work on DDEs, including investigations of 
integrability criteria, the computation of densities, generalized and 
master symmetries, and recursion operators \cite{heremanetal04}.
Notable is the work by Levi and colleagues \cite{levi78,levi97}, 
Yamilov \cite{yamilov93,yamilov94} and co-workers  
\cite{adler99,cherdantsev95,cherdantsev96,shabat91,shabat97,svinolupov91},
where the classification of DDEs (into canonical forms), integrability tests, 
and connections between integrable PDEs and DDEs are analyzed in detail. 
To a large extent, the classification and integrability testing of discrete 
equations parallels continuous equations (for reviews and 
references consult \cite{adler00,mikhailov87,mikhailov90,sokolov84}).

A wealth of information about integrable DDEs can be found in papers by 
Suris \cite{suris97a,suris97b,suris98,suris99,suris01} and his 
book \cite{suris02} in progress.  
Suris and others have shown that many lattices are closely related to the 
celebrated Toda lattice \cite{toda}, its relativistic counterpart
due to Ruijsenaars \cite{ruijsenaars90}, the KvM lattice 
\cite{kacvanmoerbeke75}, and the two-component Volterra system 
\cite{shabat91,shabat97}. 


Recently, variants of the tanh-method have been successfully applied to 
many nonlinear polynomial systems of PDEs in any number of independent 
variables \cite{db03,fan2002b,fan2003b,fan2003c,parkesetal2002}.
Baldwin {\it et al.} \cite{baldwinetalsoftwarepdes01} implemented the 
tanh- method (and also sech, cn, and sn-methods) in {\it Mathematica}. 
Liu and Li implemented the tanh-method \cite{liliuRATH,liuliRATHS} in 
{\it Maple}.

%
%

While there has been considerable work done on finding exact solutions 
to PDEs, as far as we could verify, little work is being done to 
symbolically compute exact solutions of DDEs.
In this paper we present an adaptation of the tanh-method to solve nonlinear 
polynomial differential-difference equations, which to our knowledge is novel.
Our algorithm applies to semi-discrete lattices and allows one to find closed 
form solutions that are polynomial in tanh.

Although the tanh method is easy to apply, it leads to fairly cumbersome
algebra without guarantee that a tanh solution can be found 
\cite{db03}.
We therefore present \cite{baldwinetalsoftwareddes03} the fully automated 
software package, {\em DDESpecialSolutions.m\/} in {\em Mathematica}, 
which implements the algorithm.  
Without intervention from the user, our software computes traveling wave 
solutions as polynomials in $T_{\vec{n}} = \tanh\xi_{\vec{n}},$
where $\xi_{\vec{n}} = \sum_{i=1}^Q d_{i}n_i + \sum_{i=j}^N c_jx_j + \delta.$ 
The continuous variables $x_i$ and  the discrete variables $n_j$ are 
combined with constants $c_i$ and $d_j,$ and $\delta$ is the phase constant.

For systems of DDEs involving constant parameters 
(denoted by lower-case Greek letters), the software automatically 
determines the conditions on the parameters so that the given equations 
might admit polynomial solutions involving tanh.
Obviously, since $\sech^2 \xi = 1 - \tanh^2 \xi$ our code can find solutions 
in even powers of $\sech.$ 
But our code can not find solutions involving odd powers of sech.

The paper is organized as follows:  
Section~\ref{sec:PDEs} gives a discussion of the tanh-method for PDEs with 
a worked example of a two-dimensional nonlinear PDE.
Since the tanh-method for semi-discrete lattices so closely parallels the 
tanh-method for evolutions equations, 
we believe that a discussion of the continuous case increases the 
transparency of the more complicated semi-discrete case.
In Section~\ref{sec:DDEs} we present our tanh-method for DDEs with a worked 
example of a two-dimensional Toda equation followed by almost a 
dozen additional examples.  
A description of our {\em Mathematica\/} package, {\em DDESpecialSolutions.m\/}, 
is given in Section~\ref{sec:descript}.
We discuss our results and draw some conclusions in 
Section~\ref{sec:conclusion}.  
Appendix A gives the full input and output of a test case.
%
\section{Traveling Wave Solutions of PDEs}
\label{sec:PDEs}
In this section we discuss the $\tanh$-method as it applies to 
systems of $P$ polynomial differential equations, 
\begin{equation}
  \label{originalsystem}
  \vec{\Delta} ( \vec{u}({\vec{x}}), \vec{u}'({\vec{x}}), 
  \vec{u}^{\prime\prime}({\vec{x}}), \cdots, 
  \vec{u}^{(m)} ({\vec{x}}) ) = {\bf 0}, 
\end{equation}
where the dependent variable $\vec{u}$ has $P$ components $u_i,$ 
the independent variable ${\vec{x}}$ has $N$ components $x_j,$ 
and $\vec{u}^{(m)}({\vec{x}})$ denotes the collection of mixed 
derivative terms of order $m.$  
We assume that any arbitrary coefficients parameterizing the system are 
strictly positive and denoted by lower-case Greek letters.  
To simplify the notation in the examples, 
we will write $u,v,w,\dotsc$ instead of $u_1,u_2,u_3,\dotsc$ 
and $x,y,t,\dotsc$ instead of $x_1,x_2,x_3,$ etc.
\subsection{Algorithm of the tanh-method for PDEs}
\label{sec:specialsolutionsAlgo}
%
\step{Transform the PDE into a nonlinear ODE}
\label{PDEtoODEStep}
We seek solutions in the traveling frame of reference, 
\begin{equation}
  \xi = \sum_{j=1}^N c_jx_j + \delta,
\end{equation}
where the components $c_j$ of the wave vector ${\vec{x}}$ and the phase 
$\delta$ are constants.

In the tanh method, we seek polynomial solutions expressible in 
hyperbolic tangent, $T = \tanh \xi.$  
Based on the identity $\cosh^2\xi - \sinh^2\xi = 1$,
\begin{align}
  \tanh'\xi & = \sech^2\xi = 1 - \tanh^2 \xi, \\
  \tanh''\xi & = -2 \tanh\xi + 2 \tanh^3\xi, \text{ etc.}
\end{align}
Therefore, the first and consequently all higher-order derivatives are 
polynomial in $T.$  
Thus, repeatedly applying the chain rule,
\begin{equation}
  \label{chainruletanh}
  \frac{\partial \bullet}{\partial x_j}     
      = \frac{\partial \xi}{\partial x_j} \frac{dT}{d\xi} \frac{d\bullet}{dT}
           = c_j(1-T^2)\frac{d \bullet}{dT}
\end{equation}
transforms (\ref{originalsystem}) into a coupled system of nonlinear ODEs,
\begin{equation}
 \label{legendretype}
 \vec{\Gamma}(T, \vec{U}(T), \vec{U}'(T), \dotsc ) = \vec{0},
\end{equation}
where $U_i(T)$ corresponds to $u_i(\vec{x}).$

\step{Determine the degree of the polynomial solutions}
Since we seek polynomial solutions 
\begin{equation} 
  \label{polynomialsolution}
  U_i(T) = \sum_{j=0}^{M_i} a_{ij} T^j,
\end{equation}
the leading exponents $M_i$ must be determined before the $a_{ij}$ 
can be computed.  

Substituting $U_i(T)$ into (\ref{legendretype}), the coefficients of 
every power of $T$ in every equation must vanish.
In particular, the highest degree terms must vanish.
Since the highest degree terms only depend on $T^{M_i}$ in 
(\ref{polynomialsolution}), it suffices to substitute $U_i (T) = T^{M_i}$ 
into (\ref{legendretype}). 
In the resulting polynomial system $\vec{P}(T) = 0,$ equating every 
two possible highest exponents in every component $P_i$ gives a 
linear system for determining the $M_i.$
The linear system is then solved for $M_i.$

If one or more exponents $M_i$ remain undetermined, we assign a strictly 
positive integer value to the free $M_i,$ so that every equation in 
(\ref{legendretype}) has at least two different terms with equal highest 
exponents in $T.$ 

\step{Derive the algebraic system for the coefficients $a_{ij}$}
To generate the system for the unknown coefficients $a_{ij}$ and wave 
parameters $c_j$, substitute (\ref{polynomialsolution}) 
into (\ref{legendretype}) and set the coefficients of $T^j$ to zero.
The resulting nonlinear algebraic system for the unknown $a_{ij}$ 
is parameterized by the wave parameters $c_j$ and the parameters 
in (\ref{originalsystem}), if any.

\step{Solve the nonlinear parameterized algebraic system}
The most difficult step of the method is analyzing and solving the 
nonlinear algebraic system. 
To solve the system we designed a customized, yet powerful, nonlinear solver. 

The nonlinear algebraic system is solved with the following assumptions: 
\begin{itemize}
  \item 
    all parameters (lower-case Greek letters) in (\ref{originalsystem}) 
    are strictly positive. 
    (Vanishing parameters may change the exponents $M_i$ in Step 2).
  \item 
    the coefficients of the highest power terms 
    $(a_{i \, M_i},$  $i=1,\cdots,P)$ in (\ref{polynomialsolution}) are 
    all nonzero (for consistency with Step 2); and,
  \item 
    all $c_j$ are nonzero (demanded by the nature of the solutions we seek).
\end{itemize}
The algebraic system is solved recursively, starting with the simplest 
equation, and continually back-substituting solutions.  
This process is repeated until the system is completely solved.  

To guide the recursive process, we designed functions to
(i)   factor, split, and simplify the equations; 
(ii)  sort the equations according to their complexity; 
(iii) solve the equations for sorted unknowns; 
(iv)  substitute solutions into the remaining equations; and
(v)   collect the solution branches and constraints.

This strategy is similar to what one would do by hand. 
If there are numerous parameters in the system or if it is of high degree, 
there is no guarantee that our solver will return a suitable result, 
let alone a complete result. 

\step{Build and test solutions}
Substituting the solutions from Step 4 into (\ref{polynomialsolution}) 
and reversing Step 1, one obtains the explicit solutions in the 
original variables.
It is prudent to test the solutions by substituting them into 
(\ref{originalsystem}). 
\subsection{Application of the tanh-method to a PDE}
Consider the $(2+1)$ dispersive long wave system \cite{maccari},
\begin{equation}
  \label{eq:maccari}
  \left\{
    \begin{aligned}
      u_{yt} + v_{xx} + u_xu_y + uu_{xy} & = 0, \\
      v_t + u_x + u_{xxy} + u_xv + uv_x & = 0,
    \end{aligned}
  \right.
\end{equation}
which is related to the Eckhaus system.

Applying chain rule (\ref{chainruletanh}) repeatedly to (\ref{eq:maccari}), 
we get the coupled ODEs,
\begin{equation}
  \label{maccariODE}
    \begin{gathered}
      2 c_2 T (c_3 + c_1 U) U' + c_1 c_2 (T^2 - 1) {U'}^2 + 2 c_1^2 T V'   
        \hspace*{4em} \\ \hspace*{4em} 
        + (T^2 - 1) (c_2 (c_3 + c_1 U) U'' + c_1^2 V'') = 0, \\
      c_1 (1 + 2 (3 T^2-1) c_1 c_2 + V) U' + (c_3 + c_1 U) V' 
        \hspace*{4em} \\ \hspace*{4em} 
        + c_1^2 c_2 (T^2 - 1) (6 T U'' + (T^2 - 1) U''') = 0,
    \end{gathered}
\end{equation}
where $T = \tanh(c_1x + c_2y + c_3t + \delta).$

To compute the degree of the polynomial solution(s), 
substitute $U(T) = T^{M_1}$ and $V(T) = T^{M_2}$ into (\ref{maccariODE}) and 
pull off the exponents of $T$ (see Table \ref{tbl:maccariODE}).
\noindent
\begin{table}[h]
\vspace{-0.23cm}
  $$ \begin{array}{c|c}
  \text{Term} & \text{Exponents of $T$} \\ \hline \hline 
    u_{yt} & \{ M_1 - 2, M_1 \} \\ \hline
    v_{xx} & \{ M_2 - 2, M_2 \} \\ \hline 
    u_xu_y, uu_{xy} & \{ 2 M_1, 2 M_1 - 2 \} 
  \end{array} \qquad 
  \begin{array}{c|c}
  \text{Term} & \text{Exponents of $T$} \\ \hline \hline 
    v_t & \{ M_2 - 1 \} \\ \hline
    u_x & \{ M_1 - 1 \} \\ \hline
    u_{xxy} & \{ M_1 - 3, M_1 - 1, M_1 + 1 \} \\ \hline
    u_xv, uv_x & \{ M_1 + M_2 - 1 \} 
  \end{array}
  $$
  \caption{The exponents of $T$ in (\ref{maccariODE}) after 
    substituting $U_i(T) = T^{M_i}$}
  \label{tbl:maccariODE}
\end{table}
\noindent
Removing non-dominant exponents and equating possible highest exponents, 
we find $M_1 = M_2$ (from $u_{yt}$ and $v_{xx})$ 
or $2 M_1 = M_2$ (from $u_xu_y$ or $uu_{xy}$ and $v_{xx})$ 
from the first equation.  
Then, from the second equation, we find $M_2 = 2$ 
(from $u_xv$ or $uv_x$ and $u_{xxy}).$ 
This gives us two branches,
\begin{gather}
  \label{maccariMSoln1}
  \begin{cases}
    M_1 = 1, & U(T) = a_{10} + a_{11}T, \\
    M_2 = 2, & V(T) = a_{20} + a_{21}T + a_{22}T^2,
  \end{cases} \\
  \label{maccariMSoln2}
  \begin{cases}
    M_1 = 2, & U(T) = a_{10} + a_{11}T + a_{12}T^2, \\
    M_2 = 2, & V(T) = a_{20} + a_{21}T + a_{22}T^2.
  \end{cases}
\end{gather}
For the first branch, substituting (\ref{maccariMSoln1}) into 
(\ref{maccariODE}) and equating the coefficients of $T^j$ to zero gives
\begin{equation}
  \label{pdeSystem1}
  \begin{aligned}
    c_1 (2 a_{22} c_1 + a_{11}^2 c_2) & = 0, \\ 
    c_1 (2 a_{22} c_1 + a_{11}^2 c_2) & = 0, \\ 
    a_{21} c_1^2 + a_{10} a_{11} c_1 c_2 + a_{11} c_2 c_3 & = 0, \\ 
    a_{11} c_1 (a_{22} + 2 c_1 c_2) & = 0, \\ 
    a_{11} c_1 + a_{11} a_{20} c_1 + a_{10} a_{21} c_1 - 2 a_{11} c_1^2 c_2 + 
a_{21} c_3 & = 0, \\ 
    a_{11} a_{21} c_1 + a_{10} a_{22} c_1 + a_{22} c_3 & = 0.
  \end{aligned}
\end{equation}
Similarly, for the second branch, substituting (\ref{maccariMSoln2}) into
(\ref{maccariODE}) and setting the coefficients of $T^j$ to zero yields
\begin{equation}
  \label{pdeSystem2}
  \begin{aligned}
    a_{11} a_{12} c_1 c_2 & = 0, \\ 
    a_{12}^2 c_1 c_2 & = 0, \\ 
    a_{21} c_1^2 + a_{10} a_{11} c_1 c_2 - 3 a_{11} a_{12} c_1 c_2 + a_{11} 
c_2 c_3 & = 0, \\ 
    2 a_{22} c_1^2 + a_{11}^2 c_1 c_2 + 2 a_{10} a_{12} c_1 c_2 - 2 a_{12}^2 
c_1 c_2 + 2 a_{12} c_2 c_3 & = 0, \\ 
    2 a_{22} c_1^2 + a_{11}^2 c_1 c_2 + 2 a_{10} a_{12} c_1 c_2 + 2 a_{12} c_2 
c_3 & = 0, \\ 
    c_1 (a_{12} a_{21} + a_{11} a_{22} + 2 a_{11} c_1 c_2) & = 0, \\ 
    a_{12} c_1 (a_{22} + 6 c_1 c_2) & = 0, \\ 
    a_{11} c_1 + a_{11} a_{20} c_1 + a_{10} a_{21} c_1 - 2 a_{11} c_1^2 c_2 + 
a_{21} c_3 & = 0, \\ 
    a_{12} c_1 + a_{12} a_{20} c_1 + a_{11} a_{21} c_1 + a_{10} a_{22} c_1 - 
8 a_{12} c_1^2 c_2 + a_{22} c_3 & = 0.
  \end{aligned}
\end{equation}
For the first branch, solving (\ref{pdeSystem1}) under the assumption that 
$a_{11},a_{22},c_1,c_2,$ and $c_3$ are nonzero, we find 
\begin{equation}
  \label{maccariSoln}
  \begin{gathered}
    a_{10} = -\frac{c_3}{c_1}, \qquad a_{11} = \pm 2c_1, \\
    a_{20} = 2c_1c_2 - 1, \qquad a_{21} = 0, \qquad a_{22} = -2c_1c_2.
  \end{gathered}
\end{equation}
Substituting (\ref{maccariSoln}) into (\ref{maccariMSoln1}) and returning to 
$u(x,y,t)$ and $v(x,y,t),$ we get
\begin{equation}
  \label{pdesoln}
  \begin{aligned}
    u(x,y,t) & = 
      -\frac{c_3}{c_1} \pm 2c_1 \tanh( c_1x + c_2 y + c_3 t + \delta ), \\
    v(x,y,t) & = 2c_1c_2 - 1 - 2c_1c_2\tanh^2( c_1x + c_2 y + c_3 t + \delta).
  \end{aligned}
\end{equation}
In the second branch, the equation $a_{12}^2c_1c_2 = 0$ in 
(\ref{pdeSystem2}) is inconsistent with our assumption that 
$a_{12},a_{22},c_1,c_2,$ and $c_3$ are nonzero.  
This branch does not yield a solution to (\ref{eq:maccari}).

Substituting (\ref{pdesoln}) into (\ref{eq:maccari}), we verify that our 
solution does indeed satisfy the original system.
%
%
\section{Tanh method for nonlinear DDEs}
\label{sec:DDEs}
The tanh-method can be adapted to solve nonlinear polynomial DDEs.  
Apart from slight, yet important modifications, the steps mirror those 
in Section~\ref{sec:PDEs}. 

Given is a system of $M$ polynomial DDEs, 
\begin{multline}
  \label{originalsystemDDEs}
  \vec{\Delta} ( 
    \vec{u}_{\vec{n}+\vec{p}_1}(\vec{x}), 
      \vec{u}_{\vec{n}+\vec{p}_2}(\vec{x}), \dotsc,
      \vec{u}_{\vec{n}+\vec{p}_k}(\vec{x}), \\
    \vec{u}_{\vec{n}+\vec{p}_1}'(\vec{x}), 
      \vec{u}_{\vec{n}+\vec{p}_2}'(\vec{x}), \dotsc,
      \vec{u}_{\vec{n}+\vec{p}_k}'(\vec{x}), \dotsc, \\ 
    \vec{u}_{\vec{n}+\vec{p}_1}^{(r)}(\vec{x}), 
      \vec{u}_{\vec{n}+\vec{p}_2}^{(r)}(\vec{x}), \dotsc,
      \vec{u}_{\vec{n}+\vec{p}_k}^{(r)}(\vec{x})
  ) = \vec{0},
\end{multline}
where the dependent variable $\vec{u}$ has $M$ components $u_i,$ 
the continuous variable ${\vec{x}}$ has $N$ components $x_i,$ 
the discrete variable $\vec{n}$ has $Q$ components $n_j,$
the $k$ shift vectors $\vec{p}_i \in \mathbb{Z}^Q,$ 
and $\vec{u}^{(r)}({\vec{x}})$ denotes the collection of mixed 
derivative terms of order $r.$  
We assume that any arbitrary coefficients that parameterize the system 
are strictly positive and denoted by lower-case Greek letters.  

To simplify notation in the examples, 
we use dependent variables $u,v,w,\dotsc$ instead of $u_1,u_2,u_3,\dotsc,$ 
continuous independent variables $x,y,t,\dotsc$ instead of 
$x_1,x_2,x_3,\dotsc,$ and lattice points $n,m,\dotsc,$ 
instead of $n_1,n_2,$ etc.
For example, the two-component Volterra equation \cite{suris98},
\begin{equation}
  \begin{aligned}
    \dot{u}_n &= u_n ( v_n - v_{n-1} )  \\ 
    \dot{v}_n &= v_n ( u_{n+1} - u_n ),  
  \end{aligned}
\end{equation}
has $\vec{u} = (u_1,u_2) = (u,v),$ 
$\vec{x} = x_1 = t,$ $\vec{n} = n_1 = n,$ and
$\vec{p}_1 = p_1 = -1,$  $\vec{p}_2 = p_2 = 0,$  $\vec{p}_3 = p_3 = 1.$
\subsection{Algorithm of the tanh-method for DDEs}
\step{Transform the DDE into a nonlinear DDE in $T$}
\label{step:ddetoddeinT}
We seek solutions in the traveling frame of reference,
\begin{equation}
  \xi_{\vec{n}} =  \sum_{i=1}^Q d_in_i + \sum_{j=1}^N c_jx_j + \delta
    = \vec{d} \cdot \vec{n} + \vec{c} \cdot \vec{x} + \delta,
\end{equation}
where the coefficients $c_1,c_2,\dotsc,c_N,d_1,d_2,\dotsc,d_Q$ and 
the phase $\delta$ are all constants.
The dot $(\cdot)$ denotes the Euclidean inner product.

Using the properties of hyperbolic tangent, 
$T_{\vec{n}} = \tanh\xi_{\vec{n}},$ repeatedly applying the chain rule,
\begin{equation}
  \label{ddeChainRule}
  \frac{d \bullet}{d x_j} = 
    \frac{\partial \xi_{\vec{n}}}{\partial x_j}
    \frac{dT_{\vec{n}}}{d\xi_{\vec{n}}} \frac{d\bullet}{dT_{\vec{n}}}
      = c_j (1-T_{\vec{n}}^2)\frac{d \bullet}{dT_{\vec{n}}},
\end{equation}
transforms (\ref{originalsystemDDEs}) into 
\begin{multline}
  \label{originalsystemDDEsinT}
  \vec{\Delta} (
    \vec{U}_{\vec{n}+\vec{p}_1}(T_{\vec{n}}), \dotsc,
      \vec{U}_{\vec{n}+\vec{p}_k}(T_{\vec{n}}),
    \vec{U}_{\vec{n}+\vec{p}_1}'(T_{\vec{n}}), \dotsc, 
      \vec{U}_{\vec{n}+\vec{p}_k}'(T_{\vec{n}}), \\ \dotsc,
    \vec{U}_{\vec{n}+\vec{p}_1}^{(r)}(T_{\vec{n}}), \dotsc,
      \vec{U}_{\vec{n}+\vec{p}_k}^{(r)}(T_{\vec{n}})
  ) = \vec{0}.
\end{multline}
It is important to note that for any $s\, (s=1, \dotsb, k),$
$\vec{U}_{\vec{n}+\vec{p}_s}$ is a function of $T_{\vec{n}}$ and 
not $T_{\vec{n}+\vec{p}_s}.$  
Using the identity,
\begin{equation}
  \label{plusidentity}
  \tanh(x+y) = \frac{\tanh x + \tanh y}{1 + \tanh x \tanh y},
\end{equation}
we can write 
\begin{equation}
  \label{TnpToTn}
   T_{\vec{n}+\vec{p}_s} = 
     \frac{ T_{\vec{n}} + \tanh \phi_s}{1 + 
      T_{\vec{n}} \tanh \phi_s },   
\end{equation}
where 
\begin{equation}
  \label{phi}
  \phi_s = \vec{p}_s \cdot \vec{d} 
    = p_{s1}d_1 + p_{s2}d_2 + \dotsb + p_{sQ}d_Q, 
\end{equation}
and $p_{sj}$ is the $j$-th component of shift vector $\vec{p}_s.$

\step{Determine the degree of the polynomial solutions}
Seeking solutions of the form
\begin{equation}
  \label{DDEsolnForm}
  U_{i,\vec{n}}(T_{\vec{n}}) = \sum_{j = 0}^{M_i} a_{ij} T^j_{\vec{n}},
\end{equation}
we must first compute the leading exponents $M_i.$  
As in the continuous case, we can do this by substituting only the 
leading term,
\begin{equation}
  \label{TnpMi}
  U_{i, \vec{n} + \vec{p}_s}(T_{\vec{n}}) 
    = T_{\vec{n} + \vec{p}_s}^{M_i} 
    = \left[ 
        \frac{ T_{\vec{n}} +\tanh \phi_s}{1 + T_{\vec{n}} \tanh \phi_s} 
      \right]^{M_i},
\end{equation}
with $\phi_s$ in (\ref{phi}). 

Suppose we are interested in balancing terms with shift $\vec{p}_l,$ 
then terms with shifts other than $\vec{p}_l,$ say $\vec{p}_s,$ 
will not effect the balance since $U_{i,\vec{n}+\vec{p}_s}$ can be 
interpreted as being of degree zero in $T_{\vec{n}+\vec{p}_l}.$
For instance, if $\vec{p}_l = \vec{0},$ 
then $U_{i,\vec{n}+\vec{0}}(T_{\vec{n}}) = T_{\vec{n}}^{M_i}$ 
is of degree $M_i$ in $T_{\vec{n}}$ and 
$U_{i,\vec{n}+\vec{p}_s}(T) 
 = \left[ \frac{ T_{\vec{n}} +\tanh \phi_s}{1 + T_{\vec{n}} \tanh \phi_s} 
   \right]^{M_i} $ 
is of degree zero in $T_{\vec{n}}.$  

Therefore, if we need to balance terms with shift $\vec{p}_l,$ we substitute 
\begin{equation}
  \label{highestpoweransatz}
  U_{i,\vec{n} + \vec{p}_s} = 
    \begin{cases}
      \chi_i T^{M_i}_{\vec{n}}, & s = l, \\
      \chi_i, & s \neq l,
    \end{cases}
\end{equation}
into (\ref{originalsystemDDEsinT}) and proceed as in 
Step 2 of Section \ref{sec:PDEs}.  
We then continue with the union of the solutions found for $l=1,2,\dotsc,k.$  

\step{Derive the algebraic system for the coefficients $a_{ij}$}
Substitute
\begin{equation}
  \label{DDEsolutionansatz}
   U_{i, \vec{n} + \vec{p}_s}(T_{\vec{n}}) 
    = \sum_{j=0}^{M_i} a_{ij} T_{\vec{n}+\vec{p}_s}^j
    = \sum_{j=0}^{M_i} a_{ij}
    \left[ \frac{ T_{\vec{n}} + \tanh \phi_s}{1 + 
      T_{\vec{n}} \tanh \phi_s } \right]^j,   
\end{equation}
into (\ref{originalsystemDDEsinT}), 
with $\phi_s$ in (\ref{phi}).
Applying (\ref{plusidentity}) one can split $\tanh\phi_s$ into 
powers of $\tanh d_i.$ 
While doing so, we repeatedly clear the denominators.  
The resulting nonlinear algebraic system for the unknowns $a_{ij}$ is 
parameterized by $c_1,$ $c_2,$ $\dotsc,$ $c_N,$ $\tanh(d_1),$ $\tanh(d_2),$ 
$\dotsc,$ $\tanh(d_{Q})$ and any parameters (lower-case Greek letters) in 
(\ref{originalsystemDDEs}).

\step{Solve the nonlinear parameterized system}
\label{step:solveDDEsystem}
This step is the same as in the continuous case; 
we solve the system for $a_{ij}$ in terms of the parameters
$c_1,c_2,\dotsc,c_N,\tanh(d_1),\tanh(d_2),\dotsc,\tanh(d_{Q})$ and 
any parameters (lower-case Greek letters) in (\ref{originalsystemDDEs}). 

\step{Build and test the solitary wave solutions}
Substitute the solution found in Step 4 into 
(\ref{DDEsolnForm}) and reverse Step 1.  
Then, test the solutions (in the original variables) by substitution into 
(\ref{originalsystemDDEs}).
%
%
\subsection{Example of a differential-difference equation}
To illustrate the method, we derive an exact solution 
of the $(2+1)$-dimensional Toda lattice \cite{kajiwara},
\begin{equation}
  \label{exptoda}
  \frac{\partial^2 y_n}{\partial x \partial t} = \exp{(y_{n-1} - y_n)} - 
\exp{(y_n - y_{n+1})},
\end{equation}
where $y_n(x,t)$ is the displacement from equilibrium of the $n$-th unit mass 
under an exponential decaying interaction force between nearest neighbors.  
To write (\ref{exptoda}) as a polynomial DDE, set 
\begin{equation}
  \label{undtexp}
  \frac{\partial u_n}{\partial t} = \exp(y_{n-1} - y_{n}) - 1.
\end{equation}
Then, 
\begin{equation}
  \label{expuntplus1}
  \exp(y_{n-1} - y_{n}) = \frac{\partial u_n}{\partial t} + 1
\end{equation} 
and (\ref{exptoda}) becomes
\begin{equation}
  \label{varchangederivative}
  \frac{\partial^2 y_n}{\partial x \partial t} 
    = \frac{\partial u_n}{\partial t} + 
      1 - \left(\frac{\partial u_{n+1}}{\partial t} + 1\right) 
    = \frac{\partial u_n}{\partial t} - \frac{\partial u_{n+1}}{\partial t}.
\end{equation}
Integrating (\ref{varchangederivative}) and ignoring the integration constant,
we find 
\begin{equation}
  \label{ynxequalunun1}
  \frac{\partial y_n}{\partial x} = u_n - u_{n+1}.
\end{equation}
Differentiating (\ref{undtexp}) with respect to $x$ and using 
(\ref{expuntplus1}) and (\ref{ynxequalunun1}), we compute
\begin{align}
  \frac{\partial^2 u_n}{\partial x \partial t} 
    & = \frac{\partial}{\partial x}\left(\exp(y_{n-1}-y_n) - 1\right) \\
    & = \exp(y_{n-1}-y_n)\left(\frac{\partial y_{n-1}}{\partial x} - 
\frac{\partial y_n}{\partial x}\right), \\
  & = \left(\frac{\partial u_n}{\partial t} + 1\right)[(u_{n-1} - u_n) - (u_n 
      - u_{n+1})], \\
    \label{orgtoda}
  & = \left( \frac{\partial u_n}{\partial t}+1 \right) \left( u_{n-1} - 2 u_n 
      + u_{n+1} \right).
\end{align}
So, (\ref{exptoda}) can be replaced by the polynomial equation (\ref{orgtoda}).

By repeatedly applying the chain rule (\ref{ddeChainRule}) to (\ref{orgtoda}), 
one gets
\begin{multline}
  \label{todalegendre}
  c_1c_2(1-T_n^2)\left[2T_n U_{n}'-(1-T_n^2)
  U_{n}''\right] + \\
  \left[1+c_1(1-T_n^2)U_{n}'\right] 
  \left[U_{n-1}-2U_{n}+U_{n+1}\right]=0. 
\end{multline}
where $T_n = \tanh(d_1 n + c_1 x + c_2 t + \delta).$
%

For this system, we have three shifts $p_1 = -1, p_2 = 0,$ and $p_3 = 1.$  
Substituting (\ref{highestpoweransatz}) into (\ref{todalegendre}) and pulling 
off the highest exponents, 
we find $\{M_1\}$ for shift $p_1$ and $p_3,$ and 
$\{M_1, M_1 + 1, M_1 + 2, 2M_1 + 1\}$ for shift $p_2.$ 
With only one term, neither shifts $p_1$ or $p_3$ contribute any solutions.
Equating the two highest terms from the shift $p_2$, $M_1 + 2$ and 
$2M_1 + 1,$ we find $M_1 = 1.$

Substituting (\ref{DDEsolutionansatz}) into (\ref{todalegendre}), 
clearing the denominator and setting coefficients of power terms 
in $T_n$ to zero, gives
\begin{equation}
  \label{todaalgsys}
  \begin{aligned}
    a_{11} - c_1 & = 0, \\
    c_1c_2 - \tanh^2(d_1) - a_{11} c_2 \tanh^2(d_1) & = 0,  \\
    c_1c_2 - \tanh^2(d_1) - 2 a_{11} c_2 \tanh^2(d_1)+ c_1c_2 \tanh^2(d_1) 
  & = 0.
  \end{aligned}
\end{equation}
Assuming $d_1, c_1,$ and $a_{11}$ nonzero, 
the solution of (\ref{todaalgsys}) is
\begin{equation}
  \label{todaalgsol}
    a_{10} = {\rm arbitrary},  \qquad  
    a_{11} = c_1 = \frac{\sinh^2(d_1)}{c_2}.
\end{equation}
Then, the closed form solution of (\ref{orgtoda}) is
\begin{equation}
\label{todatanhsolution}
  u_n(x,t) = a_{10} + \frac{1}{c_2}\sinh^2(d_1) \tanh\left[ d_1n + 
    \frac{\sinh^2(d_1)}{c_2} x + c_2t + \delta \right]
\end{equation}
where $a_{10}, d_1,c_2$ and $\delta$ are arbitrary.
The algorithm must be repeated if any of the parameters in 
(\ref{originalsystemDDEs}) are set to zero. 
\subsection{Further examples}
We applied the tanh algorithm to solve a variety of nonlinear lattice 
equations in $(1+1)$-dimensions.
The results are summarized in Table \ref{tbl:ddeExamples}.  
For notational simplicity, we denote 
$\frac{\partial u_n}{\partial t}$ by $\dot{u}_n.$  

The Ablowitz-Ladik equation \cite{MAandJL75,MAandJL76} 
is a discretization of the nonlinear Schr\"odinger equation.  
The Toda lattices \cite{suris97a,suris98,toda} describe vibrations in 
mass-spring lattices with an exponential interaction force.
The Volterra type equations \cite{MAandJL77,hirota00,suris98} are 
discretizations of the Korteweg-de Vries (KdV) and modified KdV equations.  
%
\vskip 0.1pt
\noindent
\begin{table}[p]
  \begin{tabular}{p{.235\textwidth}|c}
  Equation Name & Equations \\ \cline{2-2}
   Reference & Solution(s) \\ \hline
  \mbox{Ablowitz-Ladik} \mbox{Lattice} & 
    \begin{math}
      \begin{aligned}
        \dot{u}_n(t) = (\alpha + u_n v_n)(u_{n+1} + u_{n-1})- 2 \alpha u_n \\
        \dot{v}_n(t) = -(\alpha + u_n v_n (v_{n+1} + v_{n-1}) + 2\alpha v_n 
      \end{aligned}
    \end{math} \\ \cline{2-2} 
    \cite{MAandJL75,MAandJL76} & 
    \begin{math}
      \begin{aligned}
        u_n(t) & = \frac{\alpha \sinh^2(d_1)}{a_{21}} \left(\pm 1 - 
          \tanh\left[d_1 n + 2 \alpha t \sinh^2(d_1) + \delta\right] \right) \\
        v_n(t) & = a_{21} (\pm 1 + \tanh\left[d_1 n + 2 \alpha \sinh^2(d_1) t + 
          \delta\right])
      \end{aligned}
    \end{math} \\ \hline
  \mbox{Toda Lattice} & 
    \begin{math}
      \ddot{u}_n(t) = (\dot{u}_n + 1)(u_{n-1} - 2u_n + u_{n+1})
    \end{math} \\ \cline{2-2} 
    \cite{toda} & 
    \begin{math}
      u_n(t) = a_{10} \pm \sinh(d_1)\tanh[d_1n \pm \sinh(d_1)t + \delta]
    \end{math} \\ \hline
  \mbox{(2+1) Dimensional} \mbox{Toda Lattice} & 
    \begin{math}
      \frac{\partial^2 u_n}{\partial x \partial t}(x,t) = 
        \left( \frac{\partial u_n}{\partial t}+1 \right) 
        \left( u_{n-1} - 2 u_n + u_{n+1} \right) 
    \end{math} \\ \cline{2-2}
    \cite{kajiwara} & 
    \begin{math}
      u_n(x,t) = a_{10} + \frac{1}{c_2}\sinh^2(d_1) \tanh\left[ d_1n + 
        \frac{\sinh^2(d_1)}{c_2} x + c_2t + \delta \right]
    \end{math} \\ \hline
  \mbox{Another Toda} \mbox{Lattice} & 
    \begin{math}
      \begin{aligned}
        \dot{u}_n(t) & = u_n(v_n - v_{n-1}) \\
        \dot{v}_n(t) & = v_n(u_{n+1} - u_n)
      \end{aligned}
    \end{math} \\ \cline{2-2} 
    \cite{suris97a} & 
    \begin{math}
      \begin{aligned}
        u_n(t) & = -\coth(d_1)c_1 + c_1\tanh(d_1 n + c_1 t + \delta) \\
        v_n(t) & = -\coth(d_1)c_1 - c_1\tanh(d_1 n + c_1 t + \delta)
      \end{aligned}
    \end{math} \\ \hline
  \mbox{Relativistic Toda} \mbox{Lattice} & 
    \begin{math}
      \begin{aligned}
        \dot{u}_n(t) &= (1 + \alpha u_n)(v_n-v_{n-1})  \\ 
        \dot{v}_n(t) &= v_n (u_{n+1}-u_n + \alpha v_{n+1} - \alpha v_{n-1})
      \end{aligned} 
    \end{math} \\ \cline{2-2} 
    \cite{suris98} & 
    \begin{math}
      \begin{aligned}
        u_n(t) &= -\frac{1}{\alpha}  - c_1 \, {\coth}(d_1) + 
                  c_1 \, {\tanh}(d_1 n + c_1 t + \delta)  \\
        v_n(t) &= \frac{c_1}{\alpha} \, {\coth}(d_1) - 
                  \frac{c_1}{\alpha}\, {\tanh}(d_1 n + c_1 t + \delta)
      \end{aligned}
    \end{math} \\ \hline
  \mbox{Another Relativistic} \mbox{Toda Lattice} & 
    \begin{math}
      \begin{aligned}
       \dot{u}_n(t) & = (u_{n+1} - v_n)v_n - (u_{n-1} - v_{n-1})v_{n-1} \\
       \dot{v}_n(t) & = v_n(u_{n+1} - u_n)
      \end{aligned}
    \end{math} \\ \cline{2-2} 
    \cite{suris97a} &
    \begin{math}
      \begin{aligned}
        u_n(t) & = \coth(d_1)c_1 + c_1\tanh(d_1 n + c_1 t + \delta) \\
        v_n(t) & = \coth(d_1)c_1 + c_1\tanh(d_1 n + c_1 t + \delta)
      \end{aligned}
    \end{math}  \\ \hline
  \mbox{Volterra} \mbox{Lattice} & 
    \begin{math}
      \begin{aligned}
        \dot{u}_n(t) &= u_n ( v_n - v_{n-1} )  \\ 
        \dot{v}_n(t) &= v_n ( u_{n+1} - u_n )  
      \end{aligned}
    \end{math} \\ \cline{2-2} 
    \cite{suris98} & 
    \begin{math}
      \begin{aligned}
        u_n(t) &= -c_1 \,{\coth}(d_1) 
        + c_1 \, {\tanh}(d_1 n + c_1 t + \delta)  \\ 
        v_n(t) &= - c_1 \,{\coth}(d_1) 
        - c_1 \, {\tanh}(d_1 n + c_1 t + \delta) 
      \end{aligned}
    \end{math} \\ \hline
   \mbox{Discretized mKdV} \mbox{Lattice} & 
    \begin{math}
      \dot{u}_n(t) = (\alpha - u_n^2)(u_{n+1} - u_{n-1})
    \end{math} \\ \cline{2-2} 
    \cite{MAandJL77} &
    \begin{math}
      u_n(t) = \pm \sqrt{\alpha} \tanh(d_1) 
        \tanh[ d_1 n + 2 \alpha \tanh(d_1) t + \delta ]
    \end{math} \\ \hline
  \mbox{Hybrid Lattice} & 
    \begin{math}  
      \dot{u}_n(t) = (1 + \alpha u_n + \beta u_n^2)(u_{n-1} - u_{n+1})
    \end{math} \\ \cline{2-2} 
    \cite{hirota00} & 
    \begin{math}
      u_n(t) = \frac{-\alpha \pm \sqrt{\alpha^2 - 4\beta} \tanh(d_1)}{2\beta} 
  \tanh\left[d_1 n + 
        \frac{\alpha^2 - 4\beta}{2\beta} \tanh(d_1) t + \delta\right]
    \end{math}  
  \end{tabular}
  \caption{Examples of DDEs and their solutions computed with
    {\em DDESpecialSolutions.m}.}
  \label{tbl:ddeExamples}
\end{table}

\section{Description of {\em DDESpecialSolutions.m\/}}
\label{sec:descript}
The format of {\em DDESpecialSolutions\/} is similar to the {\em Mathematica\/}
function {\em DSolve.}
The output is a list of sub-lists with solutions and constraints. 
The Backus-Naur form of the function is:
\begin{eqnarray*}
\langle Main\, Function 
\rangle & \to & \mathtt{DDESpecialSolutions}[\langle Equations \rangle, 
\langle Functions \rangle, \\* 
& & \qquad \langle Discrete~Variables \rangle, 
  \langle Continuous~Variables \rangle, \\
& & \qquad \langle Parameters \rangle, \langle Options \rangle] \\ 
\langle Options \rangle & \to & 
\verb|Verbose|\rightarrow \langle Bool \rangle \; | 
\verb|InputForm|\rightarrow \langle Bool \rangle \; | \\*
&& \verb|DegreeOfThePolynomial| \rightarrow 
\langle List\, of\, Rules \rangle \; | \\*
&& \verb|NumericTest| \rightarrow \langle Bool \rangle \; 
| \; \verb|SymbolicTest| \rightarrow \langle Bool \rangle \\
\langle Bool \rangle & \to & \mathtt{True} \; | \; \mathtt{False} \\
\langle List\, of\, Rules \rangle & \to & \mathtt{
\{ m[1] \rightarrow Integer, m[2] \rightarrow Integer,... \} } 
\end{eqnarray*}

The main function, {\em DDESpecialSolutions\/}, 
uses five subroutines corresponding to the five steps of the algorithm.  
\begin{enumerate}
  \item {\em DDESpecialSolutionsVarChange\/} returns a list of 
    nonlinear DDEs in $T_{\vec{n}}.$  
  \item {\em DDESpecialSolutionsmSolver\/} uses the result from 
    {\em DDESpecialSolutionsVarChange\/} to determine the degree(s), $M_i,$
    of the polynomial solutions.  
  \item {\em DDESpecialSolutionsBuildSystem\/} uses the output of 
    {\em DDESpecialSolutionsVarChange\/} and {\em DDESpecialSolutionsmSolver\/} 
    to build the nonlinear algebraic system for the $a_{ij}.$  
    The output of {\em DDESpecialSolutionsBuildSystem\/} is a list of sub-lists 
    containing the algebraic system, the unknowns, the wave parameters 
    $(d_i,c_j),$ the parameters $(\alpha,\beta,\dotsc),$ and the 
    nonzero variables $(a_{i,M_i},d_i,c_j,\alpha,\beta,\dotsc).$
  \item {\em Algebra`AnalyzeAndSolve`AnalyzeAndSolve\/} solves the system 
    returned by {\em DDESpecialSolutionsBuildSystem\/}.
  \item {\em DDESpecialSolutionsBuildSolutions\/} builds and tests the solutions 
    using the output from {\em Algebra`AnalyzeAndSolve`AnalyzeAndSolve\/}, 
    {\em DDESpecialSolutionsmSolver\/}, and the original system of DDEs.
\end{enumerate}

\section{Discussion and Conclusions}
\label{sec:conclusion}
We presented a straightforward algorithm to compute special 
solutions of nonlinear polynomial DDEs, without using explicit integration. 
We designed the symbolic package {\em DDESpecialSolutions.m\/} 
to find traveling wave solutions of nonlinear DDEs involving 
the $\tanh$ functions. 
Our code is designed to handle DDEs with multiple continuous variables 
and/or multiple lattice points $(n,m,\cdots).$ 
However, we were unable to find semi-discrete lattices with multiple 
lattice points in the literature.

While the software reproduces the known solutions for many equations, 
there is no guarantee that the code will compute the \emph{complete} 
solution set of \emph{all} polynomial solutions involving the $\tanh$ 
function, especially when the DDEs have parameters. 
This is due to restrictions on the form of the solutions and the 
limitations of the algebraic solver. 
Furthermore, the nonlinear constraints which arise in solving the
nonlinear algebraic system may be quintic or of higher degree, and 
therefore unsolvable in analytic form.  
Also, since our software package is fully automated, it may not return the 
solutions in the simplest form. 

Often, the nonlinear solver returns constraints on the wave parameters
$c_j$ and the external parameters (lower-case Greek letters), if any. 
In principle, one should verify whether or not such constraints affect 
the results of the previous steps in the algorithm. 
In particular, one should verify the consistency with the results 
from Step 2 of the algorithm. 

There are several ways our algorithm could be generalized.  
One could look for solutions other than polynomials in tanh. 
Rational solutions would be an obvious choice since 
$T_{\vec{n}+\vec{p}_j}$ is a rational expression in $T_{\vec{n}}.$
One may try to compute exact solutions in terms of functions $f$ 
(other than tanh).
We restricted our algorithm to tanh solutions because 
$f(\xi_{\vec{n}+\vec{p}_j}) = \tanh(\xi_{\vec{n}+\vec{p}_j})$ 
can be expressed in $f(\xi_{\vec{n}}) = \tanh(\xi_{\vec{n}})$ via 
\begin{equation}
f(\xi_{\vec{n}\!+\!\vec{p}_j})= \tanh(\xi_{\vec{n}\!+\!\vec{p}_j}) 
    = \frac{\tanh(\xi_{\vec{n}}) \!+\! \tanh(\phi_j)}%
      {1 \!+\! \tanh(\xi_{\vec{n}})  \tanh(\phi_j)}
     = \frac{f(\xi_{\vec{n}}) \!+\! \tanh(\phi_j)}{1 \!+\! 
 f(\xi_{\vec{n}})\tanh(\phi_j)},
\end{equation}
which allowed us to use chainrule (\ref{ddeChainRule}). 
If tanh were replaced by another function $f$, the function must have the 
property that $f(\xi_{\vec{n}+\vec{p}_j})$ can be expressed as some function
of $f(\xi_{\vec{n}}).$ 
Otherwise a modified chain rule for $f$ would no longer work. 
%
%
\section*{Acknowledgements}
\label{acknowledgements}

This material is based upon work supported by the National Science 
Foundation (NSF) under Grants Nos.\ CCR-9901929, DMS-9732069 and DMS-9912293. 
Any opinions, findings, and conclusions or recommendations expressed in this
material are those of the authors and do not necessarily reflect the views
of NSF. 

WH thankfully acknowledges the hospitality and support of the Department of 
Applied Mathematics of the University of Stellenbosch, South Africa, 
during his sabbatical visit in Spring 2001. 
Part of the work was done at Wolfram Research, Inc., while
WH was supported by a Visiting Scholar Grant in Fall 2000. 

The authors thank P.\ Blanchard, J.\ Blevins, J.\ Heath, J.\ Milwid, 
and M.\ Porter-Peden for their help with the project.  
%
\section*{Appendix A. Test run description}

We illustrate the use of the package {\em DDESpecialSolutions.m\/} on a PC. 
Users should have access to {\em Mathematica} v.\ 3.0 or higher. 

Put the package in a directory, say myDirectory, on drive C. 
Start a {\it Mathematica} notebook session and execute the commands:
\vskip 1pt
\noindent
\begin{verbatim}
SetDirectory["c:\\myDirectory"];  (* specify directory *)

<<DDESpecialSolutions.m           (* read in the package   *)

DDESpecialSolutions[
 {D[u[n, t], t] == (1 + alpha*u[n, t])*(v[n, t] - v[n - 1, t]), 
  D[v[n, t], t] == v[n, t]*(u[n+1, t] - u[n, t] + alpha*v[n+1, t] - 
   alpha*v[n - 1, t])
 }, {u, v}, {n}, {t}, {alpha}, 
 Verbose -> False, NumericTest -> True, SymbolicTest -> True
 (*, DegreeOfThePolynomial -> {m[1] -> 1, m[2] -> 1} *)]; 
\end{verbatim}
If the DegreeOfThePolynomial
$\rightarrow \{ m[1] \rightarrow 1, m[2] \rightarrow 1 \} $ 
were specified, the code would continue with this case only and not attempt to 
compute the degrees of the polynomials $U(T_{\vec{n}})$ and $V(T_{\vec{n}}).$

The output of the above is:

\hrule

DDESpecialSolutionsmSolver::``remove'' : 
\begin{quote}
    \vspace{-1em}
    The potential solutions $\{\{ m[1]\to 1, m[2]\to 2\}\}$ are being removed 
because they are (i) negative, (ii) contain freedom, (iii) fail to balance 
highest exponent terms from two different terms in the original system. If 
$M_i < 0,$ then the transformation $u \to 1/v$ may result in a system that 
DDESpecialSolutions can solve.
\end{quote}

Numerically testing the solutions.

Symbolically testing the solutions.
\begin{multline*}
\{ \{ \{ {{u(n,t)}\rightarrow 
      {-\left( \frac{1 + \alpha\,c(2)\,\coth (c(1)) - \alpha\,c(2)\,\tanh 
(\mbox{phase} + n\,c(1) + t\,c(2))}
          {\alpha} \right) }}, \\ {{v(n,t)}\rightarrow 
      {\frac{c(2)\,\left( \coth (c(1)) - \tanh (\mbox{phase} + n\,c(1) + 
t\,c(2)) \right) }{\alpha}}}\} \} \}
\end{multline*}

\hrule

The package {\em DDESpecialSolutions.m\/} has been tested on various
PCs with {\it Mathematica} versions 3.0, 4.0, 4.1 and 5.0. 


\begin{sidewaystable}[p]
  \newlength{\gnat}
  \settowidth{\gnat}{Another Relativistic}
  \begin{tabular}{p{\gnat}|c|c|c}
  Equation Name & Equations & Solution(s) & Ref. \\ \hline
  \mbox{Ablowitz-Ladik} \mbox{Lattice} & 
    \begin{math}
      \begin{aligned}
        \dot{u}_n(t) = (\alpha \!+\! u_n v_n)(u_{n\!+\!1} \!+\! 
u_{n\!-\!1})\!-\! 2 \alpha u_n \\
        \dot{v}_n(t) = \!-\!(\alpha \!+\! u_n v_n (v_{n\!+\!1} \!+\! 
v_{n\!-\!1}) \!+\! 2\alpha v_n 
      \end{aligned}
    \end{math} & 
    \begin{math}
      \begin{aligned}
        u_n(t) & = \frac{\alpha \sinh^2(d_1)}{a_{21}} \left(\pm 1 \!-\! 
          \tanh\left[d_1 n \!+\! 2 \alpha t \sinh^2(d_1) \!+\! \delta\right] 
\right) \\
        v_n(t) & = a_{21} (\pm 1 \!+\! \tanh\left[d_1 n \!+\! 2 \alpha 
\sinh^2(d_1) t \!+\! 
          \delta\right])
      \end{aligned}
    \end{math} & 
    \cite{MAandJL75,MAandJL76} \\ \hline
  \mbox{Toda Lattice} & 
    \begin{math}
      \ddot{u}_n(t) = (\dot{u}_n \!+\! 1)(u_{n\!-\!1} \!-\! 2u_n \!+\! 
u_{n\!+\!1})
    \end{math} & 
    \begin{math}
      u_n(t) = a_{10} \pm \sinh(d_1)\tanh[d_1n \pm \sinh(d_1)t \!+\! \delta]
    \end{math} &
    \cite{toda} \\ \hline
  \mbox{(2+1) Dimensional} \mbox{Toda Lattice} & 
    \begin{math}
      \frac{\partial^2 u_n}{\partial x \partial t}(x,t) = 
        \left( \frac{\partial u_n}{\partial t}\!+\!1 \right) 
        \left( u_{n\!-\!1} \!-\! 2 u_n \!+\! u_{n\!+\!1} \right) 
    \end{math} & 
    \begin{math}
      u_n(x,t) = a_{10} \!+\! \frac{1}{c_2}\sinh^2(d_1) \tanh\left[ d_1n \!+\! 
        \frac{\sinh^2(d_1)}{c_2} x \!+\! c_2t \!+\! \delta \right]
    \end{math} & 
    \cite{kajiwara} \\ \hline
  \mbox{Another} \mbox{Toda Lattice} & 
    \begin{math}
      \begin{aligned}
        \dot{u}_n(t) & = u_n(v_n \!-\! v_{n\!-\!1}) \\
        \dot{v}_n(t) & = v_n(u_{n\!+\!1} \!-\! u_n)
      \end{aligned}
    \end{math} &
    \begin{math}
      \begin{aligned}
        u_n(t) & = \!-\!\coth(d_1)c_1 \!+\! c_1\tanh(d_1 n \!+\! c_1 t \!+\! 
\delta) \\
        v_n(t) & = \!-\!\coth(d_1)c_1 \!-\! c_1\tanh(d_1 n \!+\! c_1 t \!+\! 
\delta)
      \end{aligned}
    \end{math} & 
    \cite{suris97a} \\ \hline
  \mbox{Relativistic} \mbox{Toda Lattice} & 
    \begin{math}
      \begin{aligned}
        \dot{u}_n(t) &= (1 \!+\! \alpha u_n)(v_n\!-\!v_{n\!-\!1})  \\ 
        \dot{v}_n(t) &= v_n (u_{n\!+\!1}\!-\!u_n \!+\! \alpha v_{n\!+\!1} \!-\! 
\alpha v_{n\!-\!1})
      \end{aligned} 
    \end{math} & 
    \begin{math}
      \begin{aligned}
        u_n(t) &= \!-\!\frac{1}{\alpha}  \!-\! c_1 \, {\coth}(d_1) \!+\! 
                  c_1 \, {\tanh}(d_1 n \!+\! c_1 t \!+\! \delta)  \\
        v_n(t) &= \frac{c_1}{\alpha} \, {\coth}(d_1) \!-\! 
                  \frac{c_1}{\alpha}\, {\tanh}(d_1 n \!+\! c_1 t \!+\! \delta)
      \end{aligned}
    \end{math} & 
    \cite{suris98} \\ \hline
  \mbox{Another Relativistic} \mbox{Toda Lattice} & 
    \begin{math}
      \begin{aligned}
       \dot{u}_n(t) & = (u_{n\!+\!1} \!-\! v_n)v_n \!-\! (u_{n\!-\!1} \!-\! 
v_{n\!-\!1})v_{n\!-\!1} \\
       \dot{v}_n(t) & = v_n(u_{n\!+\!1} \!-\! u_n)
      \end{aligned}
    \end{math} &
    \begin{math}
      \begin{aligned}
        u_n(t) & = \coth(d_1)c_1 \!+\! c_1\tanh(d_1 n \!+\! c_1 t \!+\! \delta) 
\\
        v_n(t) & = \coth(d_1)c_1 \!+\! c_1\tanh(d_1 n \!+\! c_1 t \!+\! \delta)
      \end{aligned}
    \end{math} &
    \cite{suris97a} \\ \hline
  \mbox{Volterra} \mbox{Lattice} & 
    \begin{math}
      \begin{aligned}
        \dot{u}_n(t) &= u_n ( v_n \!-\! v_{n\!-\!1} )  \\ 
        \dot{v}_n(t) &= v_n ( u_{n\!+\!1} \!-\! u_n )  
      \end{aligned}
    \end{math} & 
    \begin{math}
      \begin{aligned}
        u_n(t) &= \!-\!c_1 \,{\coth}(d_1) 
        \!+\! c_1 \, {\tanh}(d_1 n \!+\! c_1 t \!+\! \delta)  \\ 
        v_n(t) &= \!-\! c_1 \,{\coth}(d_1) 
        \!-\! c_1 \, {\tanh}(d_1 n \!+\! c_1 t \!+\! \delta) 
      \end{aligned}
    \end{math} & 
    \cite{suris98} \\ \hline
   \mbox{Discretized mKdV} \mbox{Lattice} & 
    \begin{math}
      \dot{u}_n(t) = (\alpha \!-\! u_n^2)(u_{n\!+\!1} \!-\! u_{n\!-\!1})
    \end{math} &
    \begin{math}
      u_n(t) = \pm \sqrt{\alpha} \tanh(d_1) 
        \tanh[ d_1 n \!+\! 2 \alpha \tanh(d_1) t \!+\! \delta ]
    \end{math} &
    \cite{MAandJL77} \\ \hline
  \mbox{Hybrid Lattice} & 
    \begin{math}  
      \dot{u}_n(t) = (1 \!+\! \alpha u_n \!+\! \beta u_n^2)(u_{n\!-\!1} \!-\! 
u_{n\!+\!1})
    \end{math} & 
    \begin{math}
      u_n(t) = \frac{\!-\!\alpha \pm \sqrt{\alpha^2 \!-\! 4\beta} 
\tanh(d_1)}{2\beta} 
  \tanh\left[d_1 n \!+\! 
        \frac{\alpha^2 \!-\! 4\beta}{2\beta} \tanh(d_1) t \!+\! \delta\right]
    \end{math} &
    \cite{hirota00}
  \end{tabular}
  \caption{Alternate version of Table \ref{tbl:ddeExamples}.}
\end{sidewaystable}

\end{document}